\documentclass[preprint,12pt]{elsarticle}

\input{glyphtounicode}
\usepackage{amsthm}
\usepackage{amsmath,amssymb,amsfonts}%
\usepackage{multirow}%
\usepackage{tabularx}
\usepackage{caption}
\usepackage{subcaption}
\usepackage{float}
\usepackage{multicol}
\usepackage{multirow}
\usepackage{booktabs}
\usepackage[utf8]{inputenc}
\usepackage{xcolor}
\usepackage{soul}
\usepackage{listings}%
\usepackage{threeparttable}
\usepackage{textcomp}%
\usepackage{xcolor}%
\usepackage{mathrsfs}%
\usepackage[polish,english]{babel} 
\usepackage[colorlinks,citecolor=red,urlcolor=blue,bookmarks=false,hypertexnames=true]{hyperref}
\usepackage[export]{adjustbox} 

\newcommand{\Hc}{H_{\mathrm{c}}}
\newcommand{\Js}{J_{\mathrm{s}}}
\newcommand{\dw}{\delta_{\mathrm{w}}}
\newcommand{\gw}{\gamma_{\mathrm{w}}}
\newcommand{\Lex}{L_{\mathrm{ex}}}

\journal{Acta Materialia}

\begin{document}
\setlength{\emergencystretch}{3em}

\begin{frontmatter}



\title{Solidification-cell confinement of domain-wall pinning in additively manufactured ferromagnets}

\author[a]{Dennis Boakye \corref{cor}} 
\cortext[cor]{Corresponding author}
\ead{dboakye@myumanitoba.ca}
\author[b]{Eric K. K. Abavare}
\author[a]{Chuang Deng}

\affiliation[a]{organization={Mechanical Engineering, University of Manitoba},
	addressline={66 Chancellors Cir}, 
	city={Winnipeg},
	postcode={R3T 2N2}, 
	state={Manitoba},
	country={Canada}
}
\affiliation[b]{organization={Department of Physics, Kwame Nkrumah University of Science and Technology},
	city={Kumasi},
	country={Ghana}
}

\begin{abstract}
As-built printed ferromagnets typically exhibit higher coercivity than optimized wrought materials, yet existing explanations rely on empirical fits or costly simulations. Herein, we provide a missing analytical theory that links print parameters directly to cooling rates, cellular spacing, dislocation density, and domain-wall pinning coercivity. Informed by metallographic grain data and using a single fitted constant, our model predicts six experimental datasets for pure Fe, Fe-6.9Si, and a multicomponent alloy within a factor of 1.9. We demonstrate that configurational lattice distortion is negligible, implying that single-phase printed alloys follow dilute-pinning laws. Critically, we introduce a confinement factor, $E=\sqrt{\lambda_{c}/2\delta_{w}}$, proving that solidification-induced dislocation packing makes cellular microstructures harder than conventionally cold-worked metals. The framework enables an alloy-sensitivity map to screen and rank compositions before manufacturing.
\end{abstract}


\end{frontmatter}

\section{Introduction}
Additive manufacturing (AM) of soft magnet components promises topology-optimized flux paths, integrated cooling, and site-specific property control for electrical machines \cite{goll2019additive,giannotta2023review}. These abilities are not accessible to the laminate-stamping route that has dominated soft magnet manufacturing for a century \cite{chaudhary2020additive}. The resulting deficit is equally well documented: in the as-built state, printed ferromagnets are almost invariably magnetically harder than optimized wrought material \cite{kustas2022emerging}. As-built coercivities exceed annealed-wrought values by factors of three to thirty in pure iron \cite{zanni2022relationship}, silicon steels \cite{garibaldi2018effect}, Fe-Co alloys \cite{mikler2017laser,varahabhatla2024influence}, and multicomponent alloys \cite{song2023evolution}, and post-build annealing recovers only part of the deficit. Because hysteresis loss scales with the loop area \cite{grys2021attempt,luse1994discontinuous}, this coercivity penalty translates directly into machine efficiency, and the question of which microstructural feature is responsible has become one of the most frequently posed and least quantitatively answered questions in the AM magnetics literature.
	
The current quantitative tools lie at two methodological extremes. The first is that the statistical and machine-learning regressions relate laser parameters to coercivity, with useful interpolating power but no physical content or transferability between alloys \cite{ozden2024predictive}. At the other, powder-resolved multiphysics simulation chains couple melt-pool thermal fields to thermo-elasto-plasticity and micromagnetics at a computational cost of order $10^5$ core-hours per parameter set \cite{yang2023tailoring}. Even these, in the most complete study to date, report only a phenomenological correlation between coercivity and average residual stress. What is missing is the middle layer that exists for mechanical properties in the form of Kocks--Mecking state-variable theory \cite{kocks2003physics}. A closed-form, falsifiable chain from process variables to a small set of microstructural state variables to the property is needed. For yield strength, that layer is what made microstructure design quantitative. In this study we produce the corresponding layer for coercivity.
	
Two empirical developments now make a closed-form theory possible. First, the hierarchical L-PBF microstructure has been characterized thoroughly enough to be treated with state variables. These include columnar grains, a cellular solidification substructure of spacing $\lambda_c\sim0.4$--$1\,\mu$m, dislocation densities $\bar\rho\sim10^{14}\,\mathrm{m^{-2}}$, and type~II/III residual microstress \cite{lejvcek2019selective,bertsch2020origin}. More importantly, it has been shown that as-printed cellular substructures follow the same Holt/Kocks--Mecking relation between cell diameter and dislocation density as deformation-induced cell structures \cite{gallmeyer2020knowledge,holt1970dislocation}. This equivalence justifies the direct application of six decades of deformation-based magnetic pinning theory, notably including Kersten's stress theory \cite{kersten1943grundlagen}, Tr\"auble's statistical dislocation theory \cite{seeger1966moderne}, and the micromagnetic formalization of Kronm\"uller and F\"ahnle \cite{kronmuller2003micromagnetism} into solidification microstructures. To our knowledge this transplant has not previously been done, and it turns out to be not a mere relabeling. The spatial organization of AM dislocations into cell walls changes the pinning law itself (Section~\ref{sec:mod3}).
	
The paper is organized as follows. In Section~2 we develop the theory by first fixing the microstructural state variables from the experimental record. It then derives the four modules linking process to coercivity. Section 3 evaluates the theory against published magnetometry data for L-PBF pure iron and Fe-6.9Si, as these materials naturally exhibit both pinning regimes predicted by the model within a single processing method. Section 4 extends this framework to multicomponent alloys by incorporating three new physical mechanisms, including composition-dependent intrinsic constants, exchange-averaged chemical disorder, and dual-phase magnetostatic pinning. Section~5 condenses the theory into an alloy sensitivity map. Section~6 states six falsifiable predictions. Section~7 talks about the model's limitations. Detailed derivations and constants are presented in the appendices.
	
\section{Theory}

\subsection{Microstructural state variables of printed ferromagnets}\label{sec:statevar}

We start by treating a printed ferromagnet as characterized by six state variables. These consist of grain intercepts $\bar D_i$ (direction-resolved, to capture columnarity), cell spacing $\lambda_c$, mean dislocation density $\bar\rho$, microstress amplitude $\Delta\sigma$ with correlation length $L_\sigma$, secondary-phase fraction and size $(p_f, d_f)$, and porosity $(p, d_p)$. The choice of these variables is rooted in four experimental facts. 

(i) The cellular substructure is a dislocation structure. In alloys
where there is no solid-state transformation, L-PBF produces columnar grains containing
solidification cells whose walls are decorated by dense dislocation tangles
and microsegregation \cite{bertsch2020origin}. Quantitatively, as-printed Inconel
718 with cell diameter $620\pm120$\,nm exhibits
$\bar\rho=1.6\pm0.8\times10^{14}\,\mathrm{m^{-2}}$, in agreement with the
Holt relation within one standard deviation \cite{gallmeyer2020knowledge}. The as-printed
state behaves, statistically, like a heavily cold-worked one.

(ii) Allotropy can erase the substructure. Pure iron is the exception
that proves the rule where its $\alpha\!\to\!\gamma\!\to\!\alpha$ cycling during
layer-wise reheating recrystallizes the primary structure in situ. This leaves fine equiaxed grains with no cells and no columnar
texture, while retaining $\bar\rho\sim10^{14}\,\mathrm{m^{-2}}$
\cite{lejvcek2019selective,zanni2022relationship}. Pure iron is therefore a clean realization of the
diffuse defect distribution, and cell-forming alloys of the
organized one.

(iii) Stress relief separates the variables. Annealing L-PBF iron at
850\,$^\circ$C (below the transition) reduces dislocation density by roughly
an order of magnitude \cite{song2014vacuum} and relieves microstress without
changing grain size or porosity \cite{zanni2022relationship}, a natural experiment that isolates individual terms of the pinning sum.

(iv) Porosity is a red herring above $\sim$99\% density. The same
anneal removes one third of the coercivity at strictly constant density
\cite{zanni2022relationship}, which no porosity-centered account can explain.

\subsection{Module 1: Effect of processing on solidification structure}

For a moving point source of absorbed power $P$ and speed $v$ on a half-space
of thermal conductivity $\kappa$, Rosenthal's solution gives the temperature
field $T(r)$ in the frame of the source \cite{rosenthal1946theory}. Differentiating along the
solidification isotherm yields the cooling rate at the front,
\begin{equation}
	\dot T \;=\; G R \;\simeq\; \frac{2\pi \kappa\,(T_m-T_0)^2\,v}{P},
	\label{eq:rosenthal}
\end{equation}
with $G$ the thermal gradient, $R$ the solidification-front velocity, $T_m$
the liquidus and $T_0$ the baseplate temperature. Equation
\eqref{eq:rosenthal} is the workhorse of weld metallurgy and holds for L-PBF
at the order-of-magnitude level relevant here; keyhole corrections renormalize
the prefactor only. Rapid-solidification theory and L-PBF measurements then
give the intercellular spacing as \cite{kurz2023fundamentals}
\begin{equation}
	\lambda_c \;=\; a\,\dot T^{-n}, \qquad n\simeq\tfrac{1}{3},
	\label{eq:cell}
\end{equation}
with $a$ an alloy constant of order
$80\,\mu\mathrm{m\,(K/s)^{1/3}}$ \cite{katayama1984solidification,elmer1989microstructural}. Standard cooling rates of $10^5$--$10^7$\,K/s yields $\lambda_c=0.4$--$1.7\,\mu$m, bracketing all reported cell spacings. Columnar grain intercepts follow from the melt-pool
geometry and epitaxial selection. For the purposes of this work $\bar D_i$ is
taken from metallography.

\subsection{Module 2: Effect of solidification structure and defect state}

The Holt relation for as-printed material \cite{gallmeyer2020knowledge,holt1970dislocation},
fixes the mean dislocation density from the cell spacing,
\begin{equation}
	\bar\rho \;=\; \frac{200}{\pi\lambda_c^{2}},
	\label{eq:holt}
\end{equation}
giving $\bar\rho=2\times10^{13}$--$5\times10^{14}\,\mathrm{m^{-2}}$ over the
same window, in agreement with TEM and XRD estimates for as-built metals
\cite{lejvcek2019selective,bertsch2020origin}. The physical origin is the accommodation of thermal contraction mismatch between cell core and solute-enriched wall during the final stages of solidification and subsequent thermal cycling. Holt notes that Eq.~\eqref{eq:holt} is not expected to hold when dislocation mobility is high enough to permit substantial annihilation, in which case specifically structured low-energy boundaries form instead \cite{holt1970dislocation}; that the as-printed data of Ref.~\cite{gallmeyer2020knowledge} nevertheless follow the relation is an empirical result rather than a foregone one. An equivalent geometrically-necessary estimate $\bar\rho\simeq C\,\varepsilon_{\rm th}/(b\lambda_c)$ with thermal mismatch strain $\varepsilon_{\rm th}\sim10^{-2}$ reproduces the same order.

It is worth noting at this point a numerical coincidence that motivates the regime structure of Section~\ref{sec:mod3}. At $\bar\rho=1.8\times10^{14}\,\mathrm{m^{-2}}$ the mean spacing between dislocations is $1/\sqrt{\bar\rho}\simeq79$\,nm, while the $180^\circ$ wall width in iron is $\dw\simeq69$\,nm. The two lengths are set by entirely independent physics: solidification thermodynamics on one side, the competition of exchange and anisotropy on the other and yet they agree to within 13\%. Printed iron-based ferromagnets therefore sit precisely at the crossover between a wall that averages over its defect population and one that resolves individual obstacles, which is why the organization of the dislocations, and not only their number, enters the pinning law.

\begin{figure}[!ht]
	\centering
	\includegraphics[width=1\linewidth]{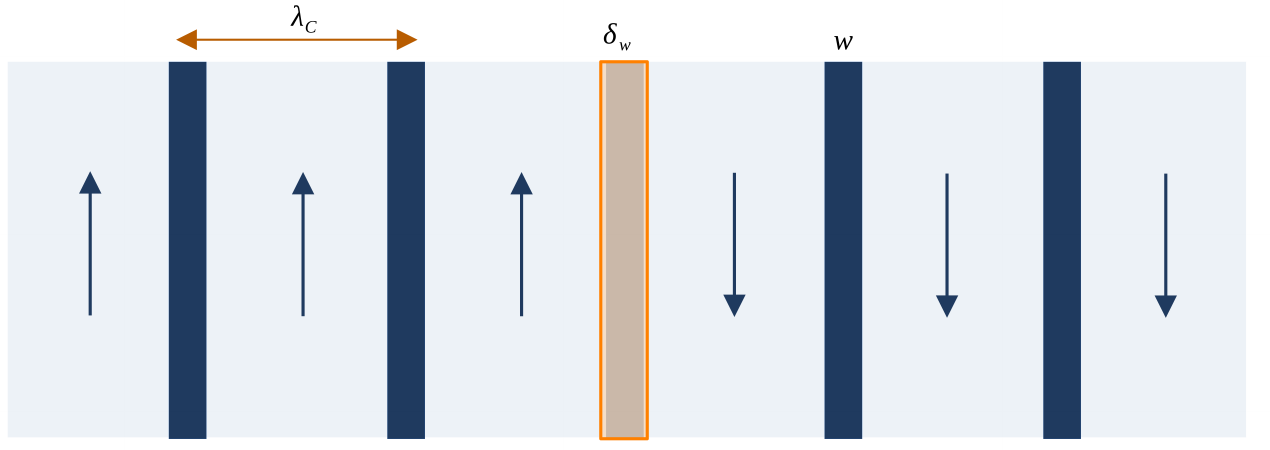}
	\caption{Geometry underlying Eqs.~\eqref{eq:partition} and \eqref{eq:confine}.
		Solidification cells of spacing $\lambda_c$ (light) separated by
		dislocation-decorated cell walls of thickness $w$ (blue), which carry the
		local density $\rho_w=\bar\rho\,\lambda_c/2w$ while the cell interiors are
		comparatively depleted. The $180^\circ$ domain wall (red) has width $\dw$ and
		translates along $x$; in iron-based alloys $\dw\simeq w$, so the domain wall
		samples the local rather than the mean dislocation density.}
	\label{fig:schematic}
\end{figure}

The dislocations are not uniformly distributed but are concentrated in cell
walls of thickness $w$ occupying volume fraction $f_w\simeq2w/\lambda_c$, so
the local wall density is (Fig.~\ref{fig:schematic})
\begin{equation}
	\rho_w \;=\; \frac{\bar\rho}{f_w} \;=\; \bar\rho\,\frac{\lambda_c}{2w}.
	\label{eq:partition}
\end{equation}

The internal stress amplitude associated with a dislocation population of
local density $\rho_{\rm loc}$ follows the Taylor hardening law
\begin{equation}
	\sigma_i=\alpha\,G_\mu b\sqrt{\rho_{\rm loc}},\qquad \alpha\simeq0.3,
	\label{eq:taylor}
\end{equation}
with $G_\mu$ the shear modulus and $b$ the magnitude of the Burgers vector. Macroscopic
(type~I) residual stress is bounded by the temperature-dependent yield stress
through the temperature-gradient mechanism. The type~II/III microstress relevant
to wall pinning has correlation length $L_\sigma\sim\lambda_c$ in cellular
alloys, analogous to grain size in cell-free metals.

\subsection{Module 3: Defect state and coercivity}\label{sec:mod3}

Consider a $180^\circ$ Bloch wall with width $\dw=\pi\sqrt{A/K_1}$ and areal energy $\gw=4\sqrt{AK_1}$, where $A$ is the exchange stiffness and $K_1$ the leading magnetocrystalline anisotropy constant. For iron, $A$ and $K_1$ of Table~\ref{tab:const} give $\dw=69$\,nm and $\gw=4.02$\,mJ/m$^2$; every number quoted below follows from this single pair, with the uncertainty propagated as stated in \ref{app:B}. Quasi-static coercivity is set by the steepest gradient of the wall's energy-per-area landscape $\gamma(x)$ along its motion coordinate,
\begin{equation}
	\Hc=\frac{1}{2\Js}\max\left|\frac{\partial\gamma}{\partial x}\right|,
	\qquad \Js=\mu_0 M_s .
	\label{eq:Hcdef}
\end{equation}
Because the defect families of Section~\ref{sec:statevar} perturb $\gamma(x)$ on
well-separated length scales, their contributions superpose to leading order
\cite{kronmuller2003micromagnetism}:
\begin{equation}
	\;\Hc \;=\; H_{\mathrm{gb}} + H_{\rho} + H_{\sigma} + H_{\mathrm{ph}} + H_{p}.\;
	\label{eq:master}
\end{equation}

Mager's bowing argument \cite{mager1952einfluss} states that a wall pinned at boundaries of mean directional intercept $\bar D_i$ must bow like a membrane of surface tension $\gw$, and depins when the Zeeman pressure exceeds the maximum restoring curvature pressure $\sim3\gw/\bar D_i$ gives
\begin{equation}
	H_{\mathrm{gb}} \;=\; \frac{3\gw}{2\Js \bar D_i}.
	\label{eq:mager}
\end{equation}
Since Eq.~\eqref{eq:Hcdef} is a maximum of a gradient and
$\max(f+g)\le\max f+\max g$, Eq.~\eqref{eq:master} is strictly an upper bound,
attained when the steepest points of the separate landscapes coincide. We use it
as an equality throughout, and the resulting bias is one-sided: the sum can only
over-predict.

For the columnar grains of most L-PBF alloys, $\bar D_i$ differs along and across the build direction, so Eq.~\eqref{eq:mager} predicts a coercivity anisotropy equal to the columnar intercept ratio.

The controlling dimensionless number of the theory is
\begin{equation}
	\eta \;=\; \frac{\lambda_c}{\dw},
\end{equation}
the ratio of the dislocation-structure scale to the domain-wall width. Two regimes follow which we discuss below.

In the diffuse regime, in which there are no cells or the defect spacing is $\lesssim\dw$, the wall overlaps many defects simultaneously. The interaction is magnetoelastic: the wall rotates the magnetization through the local Taylor stress field, picking up an energy density $\tfrac{3}{2}\tilde\lambda\,\sigma_i$, where $\tilde\lambda$ is the operative magnetostriction constant. For a wall traversing the stress field of a dislocation inside a single grain, the relevant constant is the single-crystal one associated with the shear components of that field, not the polycrystalline average; in a cubic crystal that coupling is carried by $B_2=-3\lambda_{111}c_{44}$, so for BCC iron we take $\tilde\lambda=|\lambda_{111}|=1.57\times10^{-5}$ \cite[Table I]{hall1959single}, rather than $\lambda_{100}=+2.0\times10^{-5}$ or the polycrystalline average. Published values of $\lambda_{111}$ for iron scatter by some 30\% between sources (\cite{coey2010magnetism} gives $-21\times10^{-6}$); we use \cite{hall1959single} throughout because the same campaign measured the Fe--Si series, so the whole column rests on one basis and one set of crystals. The choice of basis is load-bearing, since the anneal decrement of Section~\ref{sec:val} scales linearly with $\tilde\lambda$ and $c_1$ is fitted against it: adopting $\lambda_s$ instead would require $c_1\simeq0.75$, far outside its theoretical range, which is itself an argument that the shear constant is the operative one. Statistical averaging over defect positions dilutes the maximal gradient by a factor $c_1\simeq0.1$--0.3, and we recover the Tr\"auble form
\cite{seeger1966moderne,kronmuller2003micromagnetism}
\begin{equation}
	H_{\rho}^{\mathrm{(diffuse)}} \;=\;
	c_1\,\frac{3\tilde\lambda}{2\Js}\,\alpha G_\mu b\sqrt{\bar\rho}.
	\label{eq:trauble}
\end{equation}

The second is the confined (cellular) regime, $\eta\gg1$ with wall thickness
$w\simeq\dw$, in which the domain wall instead encounters the cell walls as discrete
planar energy wells of depth
$\Delta\gamma\simeq\tfrac{3}{2}\tilde\lambda\,\sigma_i(\rho_w)\,
\min(w,\dw)$ traversed over $\max(w,\dw)$ (\ref{app:A}). 
The magnitude of the enhancement follows in three steps from quantities already
fixed. When $w\simeq\dw$ the wall lies entirely within a single cell wall at the
steepest point of its traversal, so the stress it samples is the local
$\sigma_i(\rho_w)$ rather than the mean $\sigma_i(\bar\rho)$; the Taylor law
\eqref{eq:taylor} then converts that density ratio into a stress ratio, and the
partition \eqref{eq:partition} expresses it through measurable quantities,
\begin{equation}
	E\;=\;\frac{\sigma_i(\rho_w)}{\sigma_i(\bar\rho)}
	\;=\;\sqrt{\frac{\rho_w}{\bar\rho}}
	\;=\;\sqrt{\frac{\lambda_c}{2w}}
	\;\xrightarrow{\;w\simeq\dw\;}\;\sqrt{\frac{\lambda_c}{2\dw}} .
	\label{eq:Eheuristic}
\end{equation}
No assumption beyond $w\simeq\dw$ enters: the square root is Taylor's, and the
concentration factor is the partition. \ref{app:A} obtains the same result from
the overlap integral and supplies the behavior away from $w\simeq\dw$.
Substituting Eq. \eqref{eq:partition} converts the local quantity $\rho_w$
back to the measurable $\bar\rho$ resulting in
\begin{equation}
	H_{\rho}^{\mathrm{(cellular)}} \;=\;
	H_{\rho}^{\mathrm{(diffuse)}}\times E, \qquad
	E=\sqrt{\frac{\lambda_c}{2\dw}}=\sqrt{\frac{\eta}{2}}
	\quad (w\simeq\dw,\ \eta>2).
	\label{eq:confine}
\end{equation}
Equation \eqref{eq:confine} is the central new prediction of this paper. At equal mean dislocation density, a cellular AM microstructure is predicted to be magnetically harder than a cold-worked one by $\sqrt{\eta/2}$. This is a factor of 1.8--2.8 for $\lambda_c=0.4$--$1\,\mu$m in iron-based alloys because solidification concentrates the dislocations into sheets commensurate with the domain wall. The factor is intrinsically alloy-dependent through $\dw$: in low-anisotropy FCC compositions the wall widens, $\eta$ falls, and the enhancement fades (Section~\ref{sec:heaval}). The comparison with cold work deserves a comment, since Section~\ref{sec:statevar} rests on the equivalence of as-printed and deformation cell structures at the level of Eq.~\eqref{eq:holt}, while Eq.~\eqref{eq:confine} asserts that the two pin differently. There is no contradiction, because Eq.~\eqref{eq:holt} constrains only the cell spacing at a given $\bar\rho$ and therefore fixes $\lambda_c$ identically in the two states at matched dislocation density whereas the enhancement is governed by the wall thickness $w$ through Eq.~\eqref{eq:appAfull}. The contrast between the two states is thus carried entirely by $w$.

This makes the size of the predicted effect sensitive to a quantity that is not well constrained. Reported thicknesses of deformation cell walls in cold-worked iron span roughly 50--200\,nm, a range that straddles $\dw=69$\,nm rather than lying above it. Evaluating Eq.~\eqref{eq:appAfull} at $\bar\rho=1.6\times10^{14}\,\mathrm{m^{-2}}$ (hence $\lambda_c=0.63\,\mu$m in both states) gives a cellular-to-cold-worked ratio of 2.1 for $w\gtrsim150$\,nm, 1.8 at $w=100$\,nm, and only 1.2 at $w=50$\,nm, where the deformation structure is itself commensurate with the wall and is predicted to pin almost as strongly as the solidification structure. The theory therefore does not predict that organization per se hardens the material, but that organization commensurate with the domain wall does; whether cold work qualifies is an open experimental question rather than a settled contrast.

We regard this as sharpening P1 rather than weakening it. Because
Eq.~\eqref{eq:appAfull} gives the enhancement as a continuous function of $w/\dw$,
measuring $w$ by transmission electron microscopy alongside $\bar\rho$, $\lambda_c$
and $\Hc$ in both states tests the confinement mechanism point by point over the
accessible range; rather than as a single yes/no comparison that would return a null
result whenever the cold-worked structure happens
to sit near $w\simeq\dw$.

\begin{figure}[!ht]
	\centering
	\includegraphics[width=0.7\linewidth]{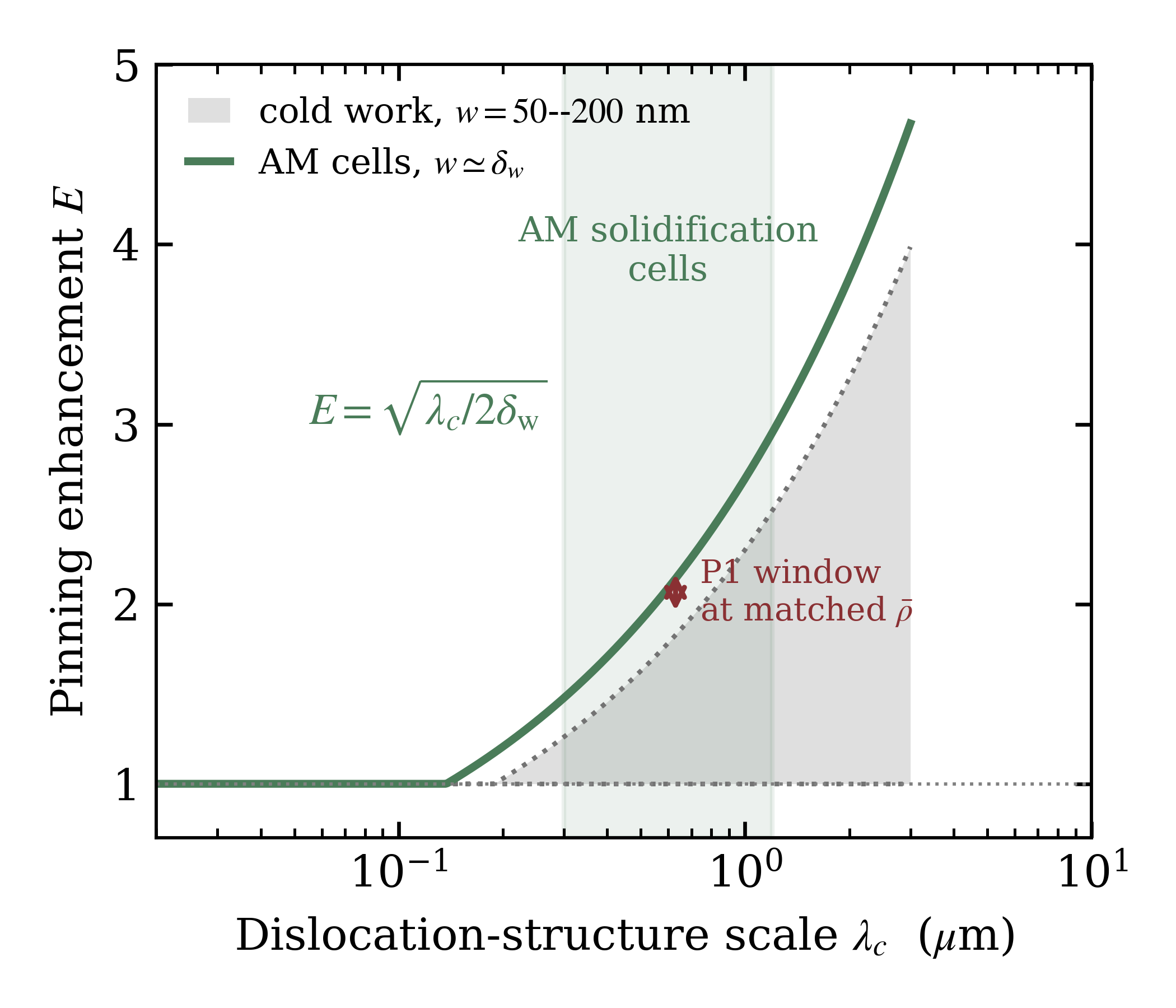}
	\caption{The confinement enhancement of Eq.~\eqref{eq:appAfull} versus the
		dislocation-structure scale, for iron ($\dw=69$\,nm). Solid: solidification
		cell walls, for which $w\simeq\dw$ and the traversal factor is unity, giving
		$E=\sqrt{\lambda_c/2\dw}$ and $E\simeq1.7$--2.7 over the L-PBF window.
		Shaded: the same expression for deformation cell walls over the reported
		thickness range $w=50$--200\,nm, where the factor $\dw/w$ suppresses the
		enhancement; both branches are floored at $E=1$, since a cellular arrangement
		cannot pin more weakly than a uniform one of the same $\bar\rho$. The
		separation between the two at matched $\bar\rho$ (arrow) is what P1 measures,
		and it collapses if the deformation walls are themselves commensurate with the
		domain wall. For low-anisotropy FCC alloys the wider wall shifts the crossover
		to larger $\lambda_c$, collapsing the enhancement (Section~\ref{sec:heaval}).}
	\label{fig:confine}
\end{figure}

A clean experimental test would require a high-magnetostriction alloy in which the dislocation term is loud enough to resolve the enhancement; we return to this in Section~\ref{sec:pred} (P1), noting here only that neither validation system of Section~3 occupies that regime. The regime structure is summarized in Fig.~\ref{fig:confine}.

Magnetoelastic pinning by stress fluctuations of amplitude $\Delta\sigma$ and correlation length $L_\sigma$ follows Kersten \cite{kersten1943grundlagen},
\begin{equation}
	H_{\sigma} \;=\; c_2\,\frac{3\tilde\lambda}{2\Js}\,
	\Delta\sigma\;\min\!\left(1,\frac{\dw}{L_\sigma}\right),
	\label{eq:stress}
\end{equation}
maximal when $L_\sigma\sim\dw$ and gradient-limited for long-wavelength stress. Type~I (macroscopic) stress shears the hysteresis loop through the magnetoelastic anisotropy but contributes little to $\Hc$. This rationalizes why full simulations find only a loose $\Hc$--$\bar\sigma$ correlation \cite{yang2023tailoring}; the mean stress is a proxy for the microstress amplitude, not the pinning agent itself.

Defects with magnetization contrast $\Delta\Js$ (pores: $\Delta\Js=\Js$; a weakly magnetic second phase: $\Delta\Js<\Js$) pin walls magnetostatically (N\'eel) and by wall-energy removal (Kersten). For a defect diameter $d$ and a volume fraction $p$,
\begin{equation}
	H_{p}\;\simeq\;c_4\,\frac{1}{2\Js}
	\left[\gw+\beta\,\frac{(\Delta\Js)^2}{\mu_0}\,\min(d,\dw)\right]
	\frac{p^{2/3}}{\max(d,\dw)},
	\label{eq:neel}
\end{equation}
Two occurrences of the domain-wall width appear here, and both express the same
restriction: a wall of width $\dw$ can neither sample an energy variation over a
distance shorter than itself, nor redistribute magnetic poles over a depth greater
than itself. The first gives the traversal distance $\max(d,\dw)$ in the
denominator; the second caps the recoverable magnetostatic energy at
$\min(d,\dw)$, since a void much larger than the wall surrenders only the pole
redistribution within a layer of thickness $\sim\dw$, not its whole demagnetizing
energy. The same argument produces the factor $\min(w,\dw)/\max(w,\dw)$ of
\ref{app:A}. Counting defects that intersect the wall at an areal density
$p^{2/3}/d^2$ and multiplying by the energy each removes then gives
Eq.~\eqref{eq:neel} directly, and the expression peaks at $d\simeq\dw$ as the
underlying inclusion theory requires \cite{kersten1943grundlagen,kronmuller2003micromagnetism};
without the two cutoffs it would instead diverge as $d\to0$ and approach a
$d$-independent constant for large defects.

The two constants are order unity and are taken from that literature rather than
fitted. $\beta$ is the geometric factor in the magnetostatic self-energy of the
pole distribution around a non-magnetic defect; for a spherical void the
demagnetizing energy density is $\tfrac12\cdot\tfrac13\,(\Delta\Js)^2/\mu_0$, so
$\beta\simeq1/6$--$1/3$, and we use $\beta=1/4$. $c_4$ is the statistical factor
converting energy per pinning site into a depinning field for a random array, the
analogue of $c_1$ in the dislocation term; Kersten-type inclusion theory places it
at 0.1--0.5 and we use $c_4=0.3$. Neither is adjusted between systems. Note that
Eq.~\eqref{eq:neel} describes discrete pinning centres interacting with a flexible
wall; at volume fractions approaching unity, or in nanograined constitutions,
individual pinning gives way to a collective regime better described by the
random-anisotropy model of Section~\ref{sec:disorder}.

Two limits need to be considered in practice, and the $\max(d,\dw)$ denominator
separates them sharply. For L-PBF gas pores ($d\sim10$--$50\,\mu\mathrm{m}\gg\dw$,
$p<1\%$), Eq.~\eqref{eq:neel} gives $H_p\simeq5$\,A/m, well under 1\% of the
as-built coercivity of printed iron and consistent with fact~(iv) of
Section~\ref{sec:statevar}: large defects are numerous in volume but sparse in the
wall plane, and each is sampled over its own diameter rather than over $\dw$.
Sub-100\,nm oxide dispersions are the opposite case. Sitting at the maximum of
Eq.~\eqref{eq:neel} they are potent out of proportion to their volume fraction,
reaching $\sim$10$^2$\,A/m at fractions of order $10^{-4}$, so that a printed alloy
with an appreciable oxide population can be pinned by it alone. The validation of
Section~\ref{sec:val} constrains the effective oxide fraction of L-PBF iron to lie
below this level, since the annealed coercivity is accounted for by the
grain-boundary floor without an additional anneal-invariant term. The same expression with $\Delta\Js$ between ferromagnetic BCC and weakly magnetic FCC supplies the dual-phase term $H_{\rm ph}$ needed for multiphase HEAs (Section~\ref{sec:dualphase}).

\subsection{Module 4: Heat treatment and coercivity recovery}

Static recovery of the dislocation network toward a pinned limit $\rho_\infty$ follows second-order annihilation kinetics,
\begin{equation}
	\rho(t,T)=\rho_\infty+\frac{\rho_0-\rho_\infty}
	{1+k(T)\,(\rho_0-\rho_\infty)\,t},
	\qquad k(T)=k_0\,e^{-Q/R_g T},
	\label{eq:recovery}
\end{equation}
with $R_g$ the gas constant (distinct from the front velocity $R$ of
Eq.~\eqref{eq:rosenthal}) and $Q$ of the order of the (pipe-)diffusion activation energy. Microstress
relaxes with the network, $\Delta\sigma\propto\sqrt{\rho}$. Above the
grain-growth threshold, parabolic growth $D^2=D_0^2+k_g t$ feeds the Mager
term. Equations \eqref{eq:master}--\eqref{eq:recovery} then return the full
$\Hc(t,T)$ master curve of a printed part through any anneal. That is a fast
defect-recovery stage saturating at the grain-boundary floor, followed by a
slow grain-growth stage that lowers the floor itself. The two stages are
cleanly separated in L-PBF iron, where an $850\,^\circ$C stress relief recovers
dislocations and microstress at fixed grain size before any growth
\cite{zanni2022relationship}, and the high-temperature grain growth of Fe-Si
follows \cite{garibaldi2018effect}. In Fe-Co the as-built$\to$annealed
softening is likewise reported \cite{varahabhatla2024influence}, though a single
combined anneal there grows the grains, recovers the defects, and develops B2
order simultaneously, so it does not by itself resolve the two stages.

\section{Experimental validation: Fe and Fe-Si}\label{sec:val}

We first test the theory against two published L-PBF systems that realize
its two regimes within a single process. Pure iron, by fact~(ii) of
Section~\ref{sec:statevar}, prints cell-free with fine equiaxed grains
($D\simeq5\,\mu$m, $\bar\rho\simeq1.6\times10^{14}\,\mathrm{m^{-2}}$)
\cite{lejvcek2019selective,zanni2022relationship} which is the diffuse limit. Fe-6.9\,wt\%\,Si, whose
$\alpha\!\to\!\gamma$ transition is suppressed, prints columnar and cellular
\cite{garibaldi2016metallurgy} giving the confined limit. Crucially, the 850\,$^\circ$C
stress relief applied to pure iron in Ref.~\cite{zanni2022relationship} changed
dislocations and stress without altering the 5\,$\mu$m grains,
isolating the terms of Eq.~\eqref{eq:master}.

\begin{figure}[!ht]
	\centering
	\includegraphics[width=0.7\linewidth]{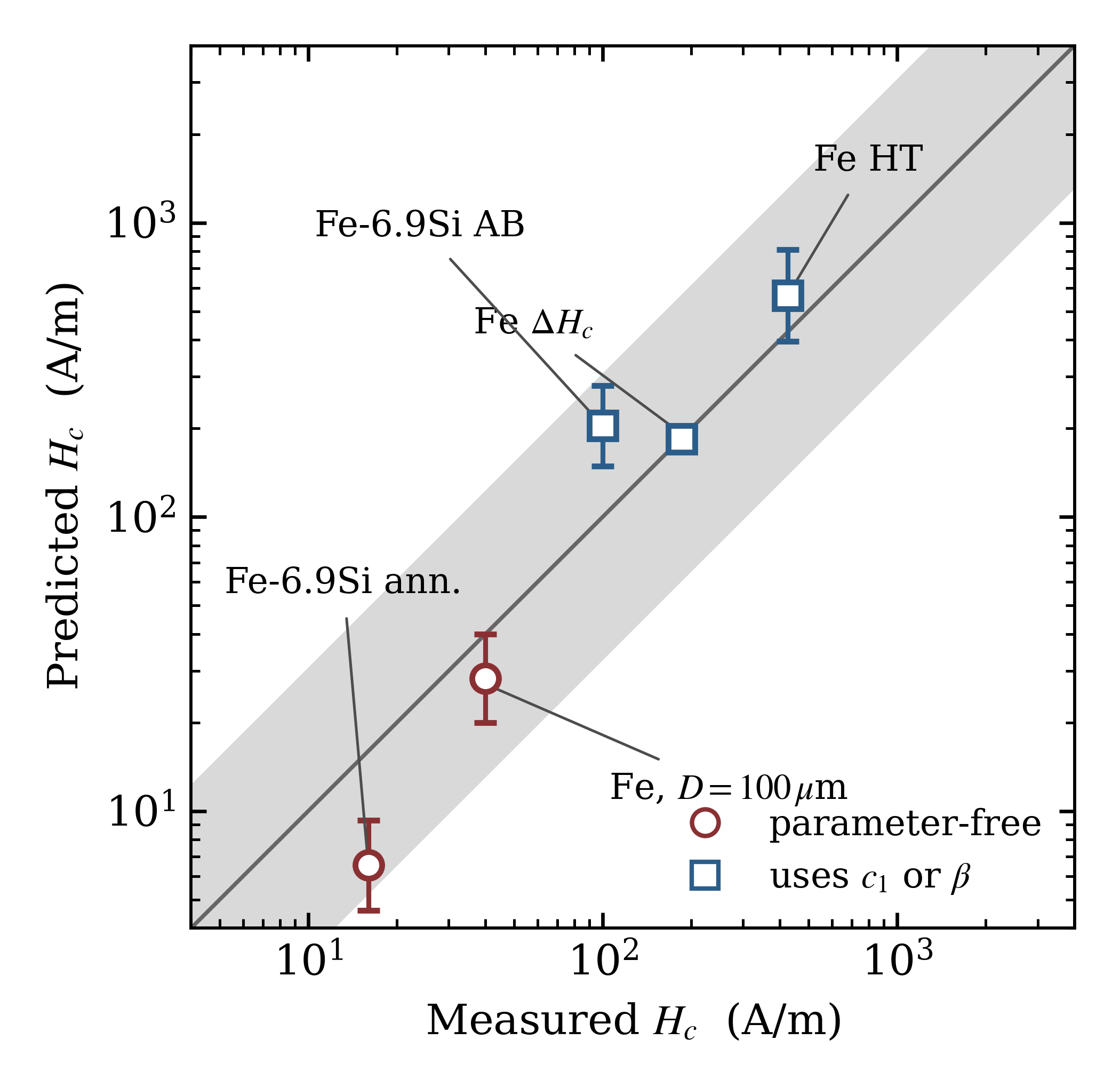}
	\caption{Parity between predicted and measured coercivity for the five
		Fe-based tests of Table~\ref{tab:val}; the band marks a factor of three.
		Circles are parameter-free predictions; the square uses the single fitted
		statistical constant $c_1=0.32$.}
	\label{fig:parity}
\end{figure}

\begin{figure}[!ht]
	\centering
	\includegraphics[width=0.7\linewidth]{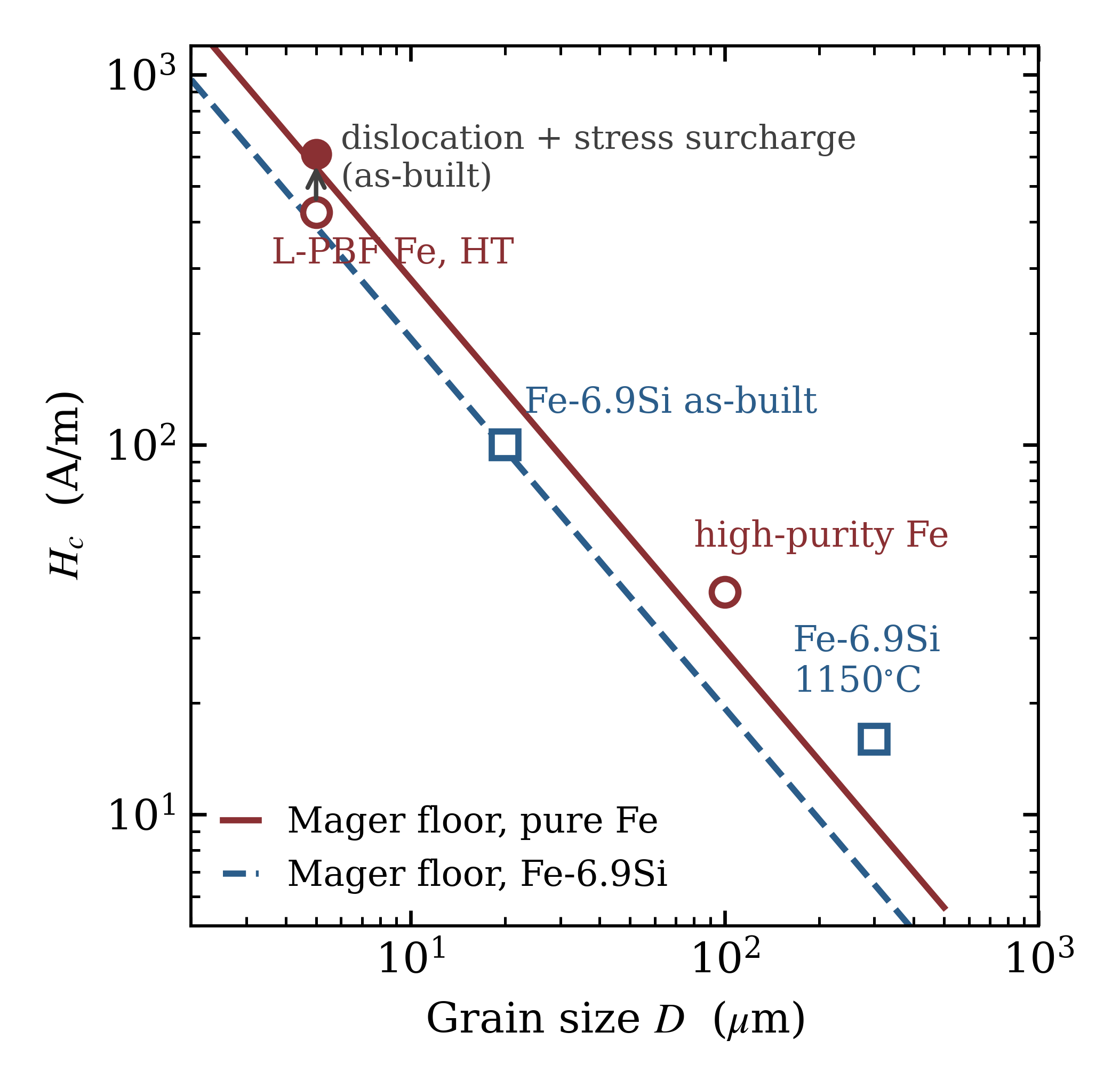}
	\caption{The Mager grain-boundary floor, Eq.~\eqref{eq:mager}, for pure Fe
		and Fe-6.9Si together with the published data. The vertical offset of
		as-built L-PBF iron above the floor is the dislocation-plus-microstress
		surcharge removed by stress relief; annealed states of both alloys track the
		floor as the grains grow.}
	\label{fig:mager}
\end{figure}

Using handbook constants only (\ref{app:B}), the model states: (i)~the stress-relieved coercivity of L-PBF iron should sit on the parameter-free Mager floor \eqref{eq:mager}; (ii)~the as-built--annealed difference should equal the Tr\"auble term \eqref{eq:trauble} evaluated between $\rho_0$ and the recovered density (the order-of-magnitude microstrain reduction measured by XRD \cite{song2014vacuum}); (iii)~as-built Fe-6.9Si, whose near-vanishing magnetostriction mutes $H_\rho$ and $H_\sigma$, should also be grain-boundary-limited at its as-built grain size of 10--30\,$\mu$m \cite{garibaldi2018effect}; and (iv)~the 1150\,$^\circ$C anneal should track the Mager floor down with the growing grains. Table~\ref{tab:val} and Fig.~\ref{fig:parity} show the outcome, while Fig.~\ref{fig:mager} displays the grain-size systematics.

\begin{table}[!ht]
	\centering\small
	\caption{Theory versus published measurement, Fe-based systems. All entries are
		computed from the single pair $(A,K_1)$ of Table~\ref{tab:const} for the
		alloy in question; no quantity is adjusted between rows. The uncertainty
		band is obtained by propagating $\pm30\%$ in the exchange stiffness
		($\pm14\%$ in $\gw$ and $\dw$) and $\pm25\%$ in the metallographic
		intercept, giving a factor $1.43/0.70$ on every grain-boundary term. Four
		rows contain no fitted quantity; the anneal decrement uses the statistical
		factor $c_1=0.32$, marginally above the range 0.1--0.3 quoted by
		\cite{kronmuller2003micromagnetism}, fitted once and then frozen for all
		subsequent systems including the HEA of Section~\ref{sec:hea}; the porosity contribution
		uses the order-unity constants $c_4$ and $\beta$ of Section~\ref{sec:mod3},
		taken from the inclusion-pinning literature rather than fitted here. The last column is the ratio of the central
		prediction to the measurement; all six lie within a factor of 2.1, and five within 1.35.}
	\label{tab:val}
	\resizebox{\textwidth}{!}{%
	\begin{tabular}{llllll}
		\toprule
		System & Quantity & Theory (central, band) & Measured & Fitted & Ratio\\
		\midrule
		L-PBF Fe \cite{zanni2022relationship} & $\Hc$, stress-relieved ($D=5\,\mu$m) & 566 (395--810)\,A/m & 425\,A/m & none & 1.33\\
		L-PBF Fe \cite{zanni2022relationship,song2014vacuum} & $\Delta\Hc$ of 850\,$^\circ$C anneal & 184 (168--202)\,A/m & 185\,A/m & $c_1$ & 1.00\\
		High-purity Fe \cite{degauque1982influence} & $\Hc$ at $D=100\,\mu$m & 28 (20--40)\,A/m & $\approx$40\,A/m & none & 0.70\\
		L-PBF Fe-6.9Si \cite{garibaldi2018effect} & $\Hc$ as-built ($D=10$--30\,$\mu$m) & 149--279\,A/m & $\approx$100\,A/m & $c_1$ & 2.04\\
		L-PBF Fe-6.9Si \cite{garibaldi2018effect} & $\Hc$ after 1150\,$^\circ$C ($D\!\approx\!300\,\mu$m) & $\gtrsim$6.5 (4.6--9.3)\,A/m & 16\,A/m & none & bound\\
		FCC HEA \cite{song2023evolution} & $\Hc$ as-built (Section~\ref{sec:heaval}) & 133--436\,A/m & 188\,A/m & $c_1$ & 1.28\\
		\bottomrule
	\end{tabular}
}
\end{table}

Four features deserve emphasis. First, the annealed L-PBF iron is grain-boundary-limited: the predicted floor is 561\,A/m from Eq.~\eqref{eq:mager} plus 5\,A/m of anneal-invariant porosity pinning, against 425\,A/m measured, a 33\% over-prediction that the propagated band of Table~\ref{tab:val} contains. The theory therefore quantifies the experimental authors' qualitative conclusion that the fine grain size, not residual defects, limits the printed material's softness \cite{zanni2022relationship}, and converts it into a design rule: softening below $\sim$100\,A/m requires $D\gtrsim50\,\mu$m, which is precisely the route the Fe-Si data follow (grains $\to300\,\mu$m, $\Hc\to16$\,A/m \cite{garibaldi2018effect}). Second, the anneal decrement closes to 4\% with one order-unity constant. Third, the porosity term is predicted, and observed, to be subdominant: the anneal removes a third of the coercivity at strictly constant density, and Eq.~\eqref{eq:neel} puts the pore contribution below 1\% of the as-built value. Fourth, the as-built Fe-6.9Si row is the only one that falls outside a factor of two, and it does so informatively. The measured 100\,A/m sits comfortably inside the Mager band of 65--194\,A/m for its 10--30\,$\mu$m grains, but the dislocation surcharge of 85\,A/m that Eq.~\eqref{eq:trauble} adds on top of it is not visible in the data. Read backwards, the measurement therefore bounds the product $\tilde\lambda\sqrt{\bar\rho}$ for this alloy below the value assumed here: either $\lambda_{111}$ continues to fall beyond the 6.37\,wt\% Si crystal from which Table~\ref{tab:const} extrapolates, or the cell spacing is coarser than the 0.5\,$\mu$m assumed. Both are checkable, and either would bring the row inside the band.

\section{Extension to concentrated multicomponent alloys}\label{sec:hea}

High-entropy and medium-entropy alloys (HEAs/MEAs) are the natural stress test
of any process--structure--property theory, because they decouple variables
that are locked together in dilute alloys: saturation magnetization,
anisotropy, magnetostriction, stacking-fault energy, and phase constitution
are all independently tunable across the composition space
\cite{chaudhary2020additive,han2022mechanically}. Soft-magnetic HEAs of the Fe-Co-Ni($X$) family have reached coercivities below 100\,A/m with gigapascal strength \cite{han2022mechanically}, and L-PBF processing of these compositions is now established \cite{song2023evolution}. To extend the theory we need three new ingredients and one re-examination.

\subsection{Intrinsic constants of concentrated solid solutions}

All structure-sensitive terms of Eq.~\eqref{eq:master} are scaled by the
intrinsic quartet $(\Js, K_1, A, \tilde\lambda)$, and in HEAs every member of
the quartet is composition-engineered rather than fixed. $\Js$ follows
Slater--Pauling-type systematics, degraded by Cr and Mn additions through
antiferromagnetic coupling; $A$ scales with the Curie temperature,
$A\propto k_B T_C/a_0$, and is therefore depressed in dilute-moment
compositions; $K_1$ in FCC Fe-Co-Ni solutions is one to two orders of
magnitude below BCC iron, which widens the domain wall
($\dw=\pi\sqrt{A/K_1}\sim150$--200\,nm); and $\tilde\lambda$ can be driven
through zero by composition. The Fe$_{0.3}$Co$_{0.5}$Ni$_{0.2}$ base of
several recent soft-magnetic HEAs was selected explicitly for minimal
saturation magnetostriction \cite{zhang2019compositional}. Within the present theory these
are not adjustable parameters but inputs, and the composition dependence of
the quartet propagates analytically into every pinning term. HEA design
already practices, implicitly, what Eq.~\eqref{eq:master} states explicitly.

\subsection{Chemical disorder: an exchange-averaging bound}\label{sec:disorder}

\begin{figure}[!ht]
	\centering
	\includegraphics[width=0.7\linewidth]{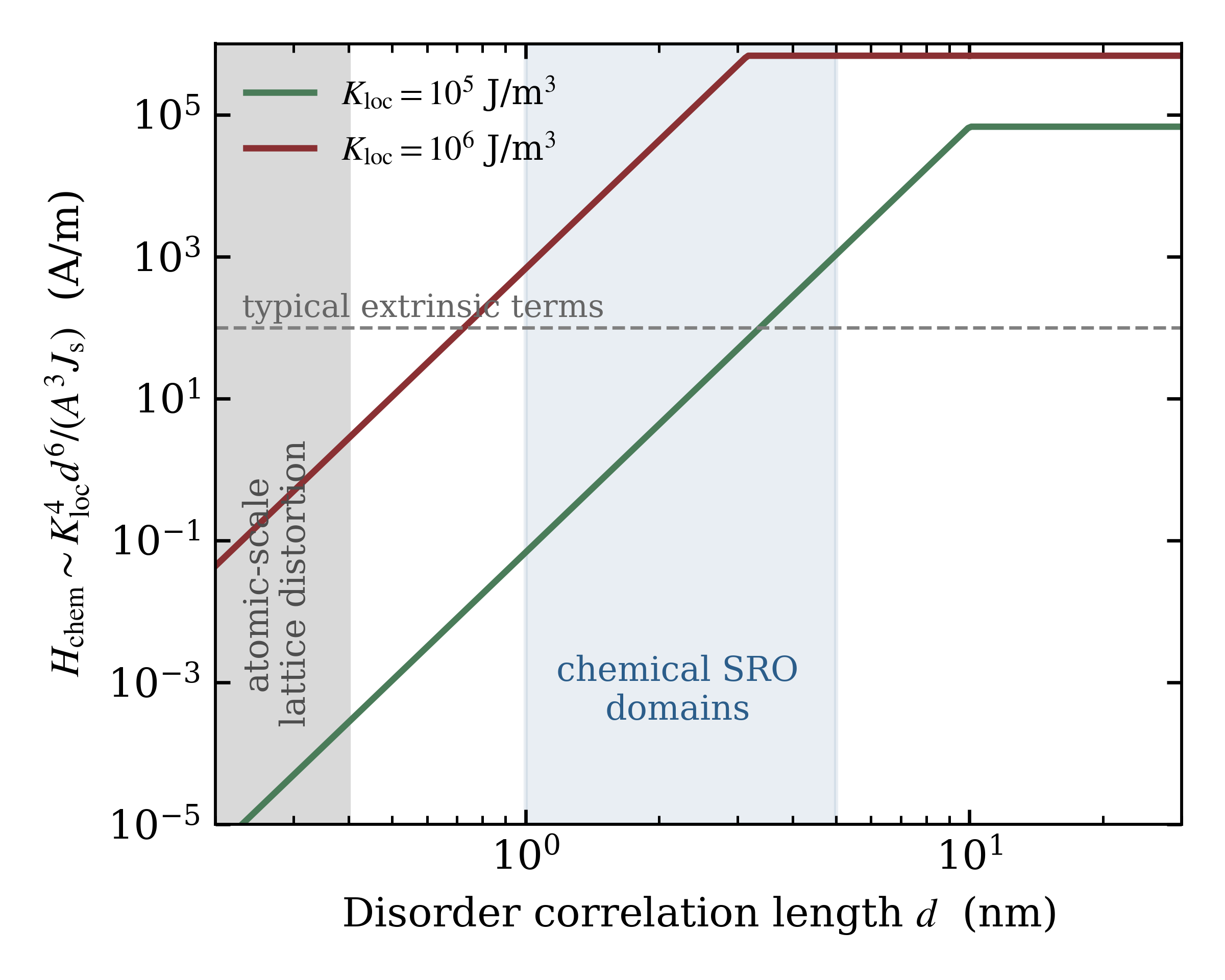}
	\caption{Exchange-averaging of chemical disorder, Eq.~\eqref{eq:herzer}.
		Atomic-scale lattice distortion (grey band) is magnetically invisible; only
		nanometre-scale chemical short-range-order domains (blue band) can approach
		the magnitude of the extrinsic pinning terms (dashed line).}
	\label{fig:herzer}
\end{figure}

The signature feature of HEAs is severe chemical disorder where every lattice site
has a different local environment, hence a fluctuating local anisotropy
$K_{\rm loc}$ and local magnetostriction. The question we ask is does this intrinsic randomness
magnetically harden the material? The random-anisotropy formalism of Alben et al. \cite{alben1978random}, developed by Herzer for nanocrystalline
magnets \cite{herzer1990grain}, answers in closed form. Fluctuations correlated
over a length $d_\chi$ (note: not defect diameter $d$ of Eq.~\eqref{eq:neel}) smaller than the exchange length
$\Lex=\sqrt{A/K_{\rm loc}}$ are averaged over $N=(\Lex/d_\chi)^3$ statistically
independent units, leaving an effective anisotropy
\begin{equation}
	\langle K\rangle \;=\; \frac{K_{\rm loc}^4\,d_\chi^6}{A^3},
	\qquad
	H_{\mathrm{chem}}\;\simeq\;\frac{p_c\,\langle K\rangle}{\Js},
	\label{eq:herzer}
\end{equation}
with $p_c$ of order unity, and valid for $d_\chi<\Lex$. The sixth-power
dependence on $d_\chi$ is devastating
for atomic-scale disorder: for $d_\chi\simeq0.3$\,nm and even a moderate
$K_{\rm loc}=10^5\,$J/m$^3$, Eq.~\eqref{eq:herzer} yields
$H_{\mathrm{chem}}\sim10^{-4}$\,A/m (Fig.~\ref{fig:herzer}). We therefore state as a result of the model that atomic-scale lattice distortion, however severe, contributes negligibly to the coercivity of a single-phase HEA and that printed HEAs are magnetically hardened by their extrinsic defect populations and phase boundaries, not by their configurational disorder. A printed single-phase HEA obeys the same
Eq.~\eqref{eq:master} as a dilute alloy, with its own intrinsic quartet.
Equation~\eqref{eq:herzer} simultaneously bounds the one disorder channel
that can matter. Chemical short-range-order (SRO) domains of $d_\chi=2$--5\,nm
with elevated $K_{\rm loc}$ rise steeply toward the extrinsic terms
(Fig.~\ref{fig:herzer}), making SRO engineering a coercivity knob unique to concentrated alloys
(Prediction P6).
	
\subsection{Sluggish diffusion and Module 4}

HEA recovery and grain-growth kinetics are slowed by the
configurational-entropy contribution to the activation barrier (``sluggish
diffusion''), entering Eq.~\eqref{eq:recovery} as an elevated effective $Q$.
The theory therefore predicts that printed HEAs retain their
dislocation-and-stress coercivity surcharge to substantially higher annealing
temperatures than dilute alloys of equal melting point, and that the two-stage
$\Hc(t,T)$ master curve of Module~4 is stretched along the $k(T)t$ axis
(Prediction P5).

\subsection{Dual-phase magnetostatic pinning}\label{sec:dualphase}

Many printable HEAs are deliberately multiphase such as BCC+B2, or mixed FCC/BCC constitutions produced by in-situ phase engineering, where ferromagnetic and weakly magnetic regions coexist on the micron scale. The magnetization contrast $\Delta\Js$ at phase boundaries makes them strong magnetostatic pinning sites through Eq.~\eqref{eq:neel} with $d\to d_f$, $p\to p_f$: for $\Delta\Js\sim1$\,T and $d_f\sim1\,\mu$m the bracket of Eq.~\eqref{eq:neel} is dominated by the magnetostatic term and $H_{\rm ph}$ reaches $\sim$$10^3$\,A/m at tens of percent phase fraction, an order of magnitude above the single-phase HEA of Section~\ref{sec:heaval}. Because the denominator is $\max(d_f,\dw)$ the contribution grows as the second phase is refined, until $d_f$ reaches $\dw$ and it saturates. In-situ phase engineering produces that second phase as ultrafine grains, which additionally raises the Mager term: Eq.~\eqref{eq:mager} at $\bar D_i\sim0.2$--$1\,\mu$m gives $H_{\rm gb}\sim3\times10^3$--$1.5\times10^4$\,A/m. The two effects act together and in the same direction, which is the experimental hierarchy reported for in-situ phase-engineered MEAs where ultrafine grains and phase interfaces are dominant, oxide particles secondary \cite{cao2024situ} and it explains why dual-phase HEA compositions sit one to two orders of magnitude above their single-phase counterparts in coercivity. The same mechanism operates at the particle scale: in an in-situ-alloyed CoFeAlMnCr HEA, in-situ Lorentz microscopy shows that Al$_2$O$_3$ ($\sim$137\,nm) and CrMn ($\sim$53\,nm) particles, both near the iron domain-wall scale where Eq.~\eqref{eq:neel} peaks, are the features that impede wall motion and set the as-built coercivity \cite{an2024achieving}, exactly as the $d\simeq\dw$ maximum of Eq.~\eqref{eq:neel} requires: evaluated at $d=137$ and 53\,nm with a 0.3\% fraction, Eq.~\eqref{eq:neel} returns $7\times10^2$ and $1.1\times10^3$\,A/m respectively, comparable to the measured as-built coercivity of that alloy.

\subsection{Validation with L-PBF HEA data}\label{sec:heaval}

The benchmark is the single-phase FCC HEA
Co$_{47.5}$Fe$_{28.5}$Ni$_{19}$Si$_{3.4}$Al$_{1.6}$ printed by L-PBF, with
optimum as-built properties $B_s=1.48$\,T, $\Hc=188$\,A/m
\cite{song2023evolution}. This is strikingly soft for an as-built part (compare 610\,A/m for
as-built iron, the sum of the stress-relieved value and the anneal decrement of
Table~\ref{tab:val}, and the vertical offset plotted in Fig.~\ref{fig:mager}), and an order of magnitude softer than the same alloy family
produced with coarser non-optimized routes ($\sim$1160\,A/m \cite{song2023evolution}).
The theory accounts for this quantitatively with the frozen $c_1=0.32$
(\ref{app:B}). The low FCC anisotropy widens the wall to $\dw\simeq180$\,nm,
which (a) collapses the confinement factor to $E\simeq1.2$ (the cellular
enhancement is an automatic casualty of low $K_1$), and (b) puts the Mager
term at only 23--70\,A/m for 10--30\,$\mu$m grains; while the
deliberately minimized magnetostriction
($\tilde\lambda\simeq3$--$10\times10^{-6}$ for the
Fe$_{0.3}$Co$_{0.5}$Ni$_{0.2}$ family \cite{zhang2019compositional}) suppresses the
dislocation term to $\sim$110--370\,A/m despite
$\bar\rho\simeq2.5\times10^{14}\,\mathrm{m^{-2}}$. The total, 133--436\,A/m,
brackets the measured 188\,A/m, and more tellingly the theory states
why this alloy prints soft when iron does not. Every one of its
intrinsic constants has been pushed in the direction that de-weights the
as-built defect population. Had $\tilde\lambda$ taken the generic transition-metal value $1.6\times10^{-5}$, the same defect state would cost
$\sim$600\,A/m. The qualitative converse is equally studied. For example Fe-Co alloys, which combine a large $\tilde\lambda$ with the highest $\Js$ of any printable ferromagnet here, are reported ``magnetically hard'' as-deposited across both L-PBF and DED until annealed \cite{mikler2017laser,varahabhatla2024influence} where they sit at the opposite corner of the sensitivity map of Section~\ref{sec:map}.

\begin{figure}[!ht]
	\centering
	\includegraphics[width=0.7\linewidth]{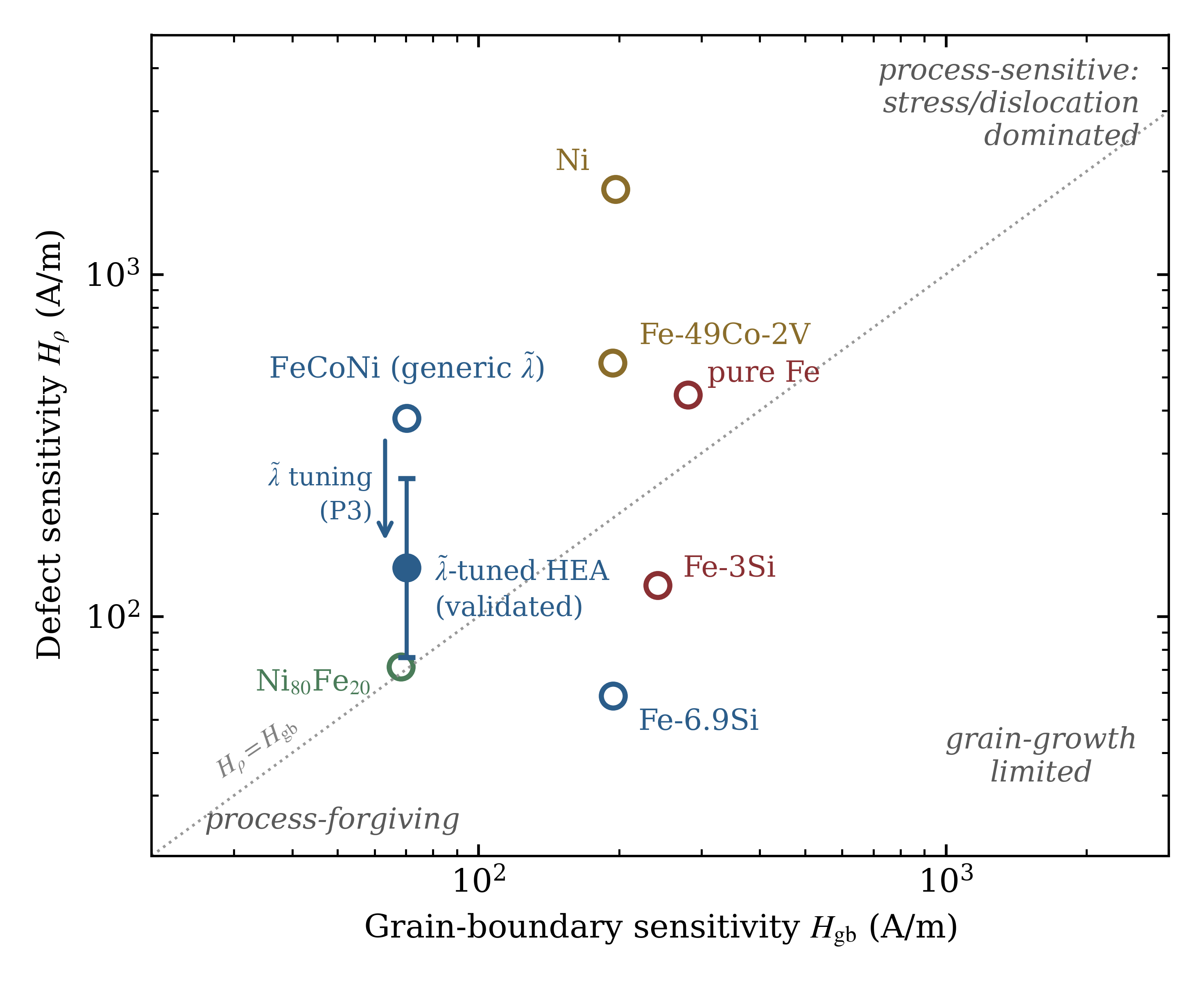}
	\caption{The alloy sensitivity map. Abscissa: grain-boundary sensitivity (Mager
		term at $D=10\,\mu$m); ordinate: defect sensitivity (dislocation term at
		$\bar\rho=10^{14}\,\mathrm{m^{-2}}$, including each alloy's own confinement
		factor). All coordinates are computed from the intrinsic quartet of
		Table~\ref{tab:const}. The arrow marks the magnetostriction-tuning trajectory
		exploited by soft-magnetic HEA design (P3); the bar spans the reported
		$\tilde\lambda=3$--$10\times10^{-6}$ of the validated composition. Its lower
		end falls essentially on nominal Ni$_{80}$Fe$_{20}$: $\tilde\lambda$-tuning
		brings a printable FCC HEA to permalloy-like process sensitivity while
		retaining a saturation polarization some 40\% higher. Fe-49Co-2V is plotted
		in its $S\!=\!0$ state, which is the one rapid solidification produces;
		developing B2 order moves it to $(57,310)$\,A/m---down and to the left, into
		the process-forgiving direction---because $\gw$ falls with $K_1$ and the
		wall widens past the cell spacing so that $E\to1$.}
	\label{fig:map}
\end{figure}

\section{The alloy sensitivity map}\label{sec:map}

Because every term of Eq.~\eqref{eq:master} factorizes into an intrinsic
prefactor times a microstructural state variable, the theory condenses into a
two-coordinate classification computed from handbook constants alone
(Fig.~\ref{fig:map}). The grain-boundary sensitivity
$3\gw/2\Js\bar D_i$ evaluated at a reference $D=10\,\mu$m, and the defect
sensitivity $c_1(3\tilde\lambda/2\Js)\alpha G_\mu b\sqrt{\bar\rho}\,E$
evaluated at a reference $\bar\rho=10^{14}\,\mathrm{m^{-2}}$. The map sorts
printable ferromagnets into three regions with distinct process strategies.
Process-forgiving compositions (permalloy; magnetostriction-minimized
FCC HEAs after $\tilde\lambda$-tuning) print soft as-built and need no
magnetic anneal. This is consistent with the simulation result that permalloy
hysteresis is insensitive to beam power \cite{yang2023tailoring}.
Grain-growth-limited compositions (high-Si steels; L-PBF iron) shed
their defect surcharge in a low-temperature stress relief but remain pinned
on the Mager floor until a high-temperature anneal grows the grains, the
two-stage behavior documented in Refs.~\cite{zanni2022relationship,garibaldi2018effect}.
Stress/dislocation-dominated compositions (Ni, and to a lesser extent Fe-Co)
are hostage to the as-built defect state and reward in-process mitigation
(preheating, scan strategy) over post-processing. Fe-49Co-2V sits in this
group on the strength of its magnetostriction, but less emphatically than a
polycrystalline $\lambda_s$ would suggest: on the resolved $\lambda_{111}$
basis used throughout (\ref{app:B}) its defect sensitivity is
$5.5\times10^2$\,A/m rather than $1.4\times10^3$, placing it near pure iron
rather than near nickel. The map is, to our knowledge, the first
instrument that ranks candidate AM magnet alloys before any printing,
and its coordinates are falsifiable one alloy at a time.

\section{Falsifiable predictions}\label{sec:pred}

\textbf{P1 (confinement).} At equal $\bar\rho$ and grain size, a cellular
as-built BCC microstructure exhibits $H_\rho$ larger than its cold-worked
counterpart by the ratio of the two $E$ factors of Eq.~\eqref{eq:appAfull}. That
ratio reaches $\sqrt{\lambda_c/2\dw}$ when the deformation walls are coarse
($w\gtrsim2\dw$) and falls towards unity when they are themselves commensurate
with the domain wall (Fig.~\ref{fig:confine}); over the reported 50--200\,nm range
of deformation cell-wall thickness it spans 1.2--2.1 in iron. The prediction is
thus quantitative in $w/\dw$ rather than a single contrast ratio, and $w$ must be
measured in both states rather than assumed. This is the
central prediction of the theory and the one that remains untested, because
the validation systems of Section~3 lie outside its window: pure iron prints
cell-free (diffuse limit, $E=1$), and Fe-6.9Si has near-vanishing
$\tilde\lambda$ that mutes $H_\rho$ regardless of organization. The clean
discriminating system is one that can be realized in both cellular and diffuse
states at matched $\bar\rho$ while retaining enough magnetostriction for
$H_\rho$ to be visible. Fe-Co is the obvious candidate, not because its
magnetostriction is the largest here---on the resolved $\lambda_{111}$ basis
it is comparable to nickel---but because it is the only alloy in
Table~\ref{tab:const} that can be cold-worked in a retained-disordered
condition and printed cellular at the same composition, which is what
matching $\bar\rho$ across the two states requires. Its
anomalously high as-deposited coercivity is already documented
\cite{mikler2017laser,varahabhatla2024influence} (as-built L-PBF Fe-49Co-2V
reaches $\approx2.1\times10^3$\,A/m, the hardest as-built state considered
here), but a decisive test demands the dislocation density, cell spacing and cell-wall
thickness of the two states, none of which the existing literature reports
alongside $\Hc$. A single campaign combining XRD line profiling (for $\bar\rho$),
transmission electron microscopy (for $\lambda_c$ and $w$) and magnetometry, on
as-built and cold-worked Fe-Co of matched $\bar\rho$, would test
Eq.~\eqref{eq:appAfull} across the accessible range of $w/\dw$ rather than at a
single point.\\
\textbf{P2 (anisotropy).} Columnar grains make $H_{\mathrm{gb}}$ directional: the
coercivity ratio along/across the build direction equals the inverse
intercept ratio of Eq.~\eqref{eq:mager}, testable on existing textured Fe-Si
builds. \\
\textbf{P3 (magnetostriction-gated process sensitivity).} Because
$H_\rho$ and $H_\sigma$ both carry $\tilde\lambda$, coercivity sensitivity to
beam power and residual stress should scale with $|\tilde\lambda|$ across
alloys and vanish at zero-magnetostriction compositions. Powder-resolved
permalloy simulations independently exhibit exactly this limit \cite{yang2023tailoring},
and the L-PBF softness of $\tilde\lambda$-minimized HEAs \cite{song2023evolution}
constitutes the complementary data point. \\
\textbf{P4 (anneal master curve).}
Equations \eqref{eq:trauble} and \eqref{eq:recovery} collapse $\Hc(t,T)$ onto
a single curve in $k(T)\,t$; the 30\% recovery at 850\,$^\circ$C/1\,h in iron
\cite{zanni2022relationship} and the sub-1000\,A/m recovery of FeSi6.5 at
1100\,$^\circ$C/1\,h fix $k_0$ and $Q$, after which the remainder of the
published anneal matrix contains no freedom.\\
 \textbf{P5 (sluggish recovery).}
For HEAs the same master curve is stretched by the elevated effective $Q$, so that
at homologous anneal temperatures, a printed HEA retains a larger fraction of
its as-built coercivity than a dilute alloy. \\
 \textbf{P6 (SRO knob).} Anneals that develop chemical
short-range order in concentrated alloys should raise $H_{\mathrm{chem}}$
with the sixth power of the SRO domain size, Eq.~\eqref{eq:herzer},
detectable as a coercivity increase during low-temperature aging of an
otherwise fully recovered single-phase HEA. The $d_\chi^6$ law applies only
while $d_\chi<\Lex=\sqrt{A/K_{\rm loc}}$; for $K_{\rm loc}=10^6\,$J/m$^3$ this
exchange length is already $\simeq3$\,nm, so the upper end of the SRO size range
saturates at $\langle K\rangle\to K_{\rm loc}$ and the predicted rise flattens
(Fig.~\ref{fig:herzer}).

\section{Limitations}
The theory is developed as a state-variable mean-field model. It ignores texture-induced anisotropy-energy variation (relevant to Fe-Si, where Goss-like fibre components reduce the effective $K_1$ and improve the annealed floor beyond the isotropic Mager estimate); it treats the additivity of Eq.~\eqref{eq:master} as exact rather than leading-order; it inherits a $\pm30\%$ uncertainty in the exchange stiffness $A$, which propagates as $\pm14\%$ to $\gw$ and $\dw$ through their square-root dependence; and the confinement factor \eqref{eq:confine} assumes $w\simeq\dw$, degrading as $w/\dw$ for thinner walls and as $\dw/w$ for coarser ones. The wall thickness $w$ is not measured in any of the systems considered here, so this assumption is the least constrained input of the framework. The HEA extension adds its own caveats: the intrinsic quartet of a given composition may be uncertain to tens of percent (especially $\tilde\lambda$ and $A$), magnetic inhomogeneity of concentrated solutions blurs the sharp $\Delta\Js$ of the dual-phase term, and the sluggish-diffusion correction to $Q$ is empirical. Chemically ordering alloys require one amendment and carry one open residual. The amendment is that $\gw=4\sqrt{AK_1}$ is a property of the ordering state, not of the alloy: in B2 Fe-Co the anisotropy falls by roughly an order of magnitude on ordering, so the wall energy, the Mager floor and the confinement factor of Eq.~\eqref{eq:confine} are all rescaled together. Table~\ref{tab:const} therefore lists Fe-49Co-2V twice, and the ordered state has $\dw=323$\,nm, wide enough that $E\to1$ and cellular confinement switches off altogether. The residual is that annealed L-PBF Fe-49Co-2V sits about an order of magnitude above the coercivity--grain-size line measured on wrought material of the same composition and ordering state \cite{yu1999pinning}, at matched grain size. That excess is grain-size independent, but it is not an antiphase-boundary pinning term: evaluated at the measured ordered-state $K_1$, such a term is some eight orders of magnitude too small, and it is bounded experimentally by the intercept of the wrought data. The excess instead tracks as-built kernel-average misorientation at fixed grain size and fixed $M_s$, and is carried by retained defect and second-phase content that survives recrystallization. The framework as it stands therefore predicts the as-built hardness of Fe-Co and the direction of its annealing response, but should not be read as walking annealed AM material down to the wrought floor. None of these caveats is structural: every term is measurable, every prediction in Section~\ref{sec:pred} can be falsified with one synchrotron or one magnetometer campaign on existing material, and the constants are few --- $c_1$ and $c_2$ are of order unity with stated ranges, while $c_4=0.3$ and $\beta=1/4$ in Eq.~\eqref{eq:neel} are taken from the inclusion-pinning literature, each carrying a factor-of-two uncertainty, and are used unchanged for pores, oxide dispersions and phase boundaries alike.

\section{Conclusion}

The Kocks--Mecking framework taught metallurgy that strength is a property of a small set of microstructural state variables with closed-form evolution laws. The present theory extends that lesson to coercivity in additive manufacturing. A printed ferromagnet's magnetic hardness is, to factor-of-two accuracy, the sum of a grain-boundary floor, a dislocation term amplified by solidification-cell confinement, a magnetostriction-gated stress term, and (in multiphase alloys) a magnetostatic phase-boundary term, each computable from build parameters and handbook constants. The extension to high-entropy alloys shows the framework at full stretch where the so-called ``lattice distortion'' is shown to be magnetically irrelevant by exchange averaging. Their softness or hardness as-printed is traced to the deliberate engineering of the intrinsic quartet, and their slow kinetics and SRO susceptibility furnish two predictions unique to the concentrated-alloy regime. The sensitivity map turns the whole construction into a design instrument such that before the first powder is melted, it tells an alloy developer whether a candidate composition will forgive the printer, demand a grain-growth anneal, or fight the residual-stress state, and it is wrong in a checkable way if the theory is.

\appendix

\section{Derivation of the confinement factor} \label{app:A}

Let the domain wall, of width $\dw$ and unperturbed energy $\gw$, move along $x$ through a periodic array of planar cell walls of spacing $\lambda_c$ and thickness $w$, each carrying local dislocation density $\rho_w$ with Taylor stress $\sigma_w=\alpha G_\mu b\sqrt{\rho_w}$. The magnetoelastic energy density acquired by wall material inside a slab is $\epsilon=\tfrac{3}{2}\tilde\lambda\,\sigma_w$ where the sign is immaterial for the gradient. The wall's excess energy as a function of its center position
$x_0$ is the overlap integral
$\Delta\gamma(x_0)=\epsilon\int\phi(x-x_0)\,\chi_w(x)\,dx$, where $\phi$ is
the normalized wall profile of support $\sim\dw$ and $\chi_w$ the slab
indicator. The overlap rises from zero to its plateau
$\epsilon\,\min(w,\dw)$ over a traversal distance $\max(w,\dw)$, so
\begin{equation}
	\max\left|\frac{\partial(\Delta\gamma)}{\partial x_0}\right|
	\;\simeq\;\frac{3}{2}\tilde\lambda\,\alpha G_\mu b\sqrt{\rho_w}\;
	\frac{\min(w,\dw)}{\max(w,\dw)} .
\end{equation}
Inserting this into Eq.~\eqref{eq:Hcdef}, eliminating $\rho_w$ via the
partition $\rho_w=\bar\rho\,\lambda_c/2w$, and absorbing the same statistical
factor $c_1$ that appears in the diffuse limit (the slabs are not perfectly
periodic or planar) gives
\begin{equation}
	H_\rho^{\mathrm{(cellular)}}=c_1\,\frac{3\tilde\lambda}{2\Js}\,
	\alpha G_\mu b\sqrt{\bar\rho}\;\sqrt{\frac{\lambda_c}{2w}}\;
	\frac{\min(w,\dw)}{\max(w,\dw)} ,
	\label{eq:appAfull}
\end{equation}
which for the regime $w\simeq\dw$ assumed throughout reduces to
Eq.~\eqref{eq:confine} with $E=\sqrt{\lambda_c/2\dw}$. The product
$\sqrt{\lambda_c/2w}\,\min(w,\dw)/\max(w,\dw)$ is maximal at $w=\dw$ and falls
away on both sides: for $w\ll\dw$ it is reduced by the factor $w/\dw$, and for
$w\gg\dw$ (coarse recovered walls) by $\dw/w$. In both limits the organization
of the dislocations ceases to be commensurate with the wall and the enhancement
is lost. The expression is not valid below unity because a cellular arrangement
cannot pin less strongly than a uniform one of the same $\bar\rho$ so
Eq.~\eqref{eq:confine} is to be read with the floor
$E=\max\!\left[1,\sqrt{\lambda_c/2\dw}\right]$, equivalent to the restriction
$\eta>2$ stated there and applied in Fig.~\ref{fig:confine}.

\section{Constants and validation arithmetic} \label{app:B}

\begin{table}[!ht]
	\centering\small
	\caption{Intrinsic constants used, with provenance. The quartet
		$(\Js,K_1,A,\tilde\lambda)$ is input; $\dw$ and $\gw$ are computed from $A$
		and $K_1$ and are not independent entries. $E$ is evaluated at
		$\lambda_c=0.6\,\mu$m; $G_\mu=81$\,GPa and $b=0.25$\,nm throughout.
		\emph{Basis of $\tilde\lambda$:} the operative constant is
		$|\lambda_{111}|$, not the polycrystalline average $\lambda_s$, because a
		wall crossing a dislocation stress field inside one grain couples to its
		shear components through $B_2=-3\lambda_{111}c_{44}$ (Section~\ref{sec:mod3}).
		Values for Fe, Fe-3Si and Fe-6.9Si are $\lambda_{111}$ measured on single
		crystals in one campaign \cite[Table I]{hall1959single}, taken in the
		more-nearly disordered state, which is the relevant one for rapid
		solidification. One entry remains an exception and is marked $\dagger$:
		for Ni$_{80}$Fe$_{20}$ a resolved $\lambda_{111}$ was not available and
		$\lambda_s$ is used instead. It does not enter Table~\ref{tab:val} and
		appears only in Fig.~\ref{fig:map}. For Fe-49Co-2V we use the
		resolved $\lambda_{111}=2.5\times10^{-5}$
		\cite[Table 4]{sundar2005soft}, consistent with
		$\lambda_{111}\simeq30\times10^{-6}$ at 45\,wt\%\,Co
		\cite[Fig.~6]{hall1959single}, in place of the polycrystalline
		$\lambda_s\simeq6.0$--$6.5\times10^{-5}$. The distinction
		is unusually large in this alloy, where $\lambda_{100}=150\times10^{-6}$
		against $\lambda_{111}=25\times10^{-6}$ \cite[Table 4]{sundar2005soft},
		so the choice of basis moves $H_\rho$ by a factor 2.6; we take
		$\lambda_{111}$ for the same reason set out in Section~\ref{sec:mod3} for
		iron. Fe-49Co-2V is listed twice because it is the one chemically ordering
		alloy here, and $\gw=4\sqrt{AK_1}$ is a property of the ordering state
		rather than of the alloy. The $S\!=\!0.88$ row is obtained from the
		measured coercivity--grain-size slope of wrought Fe-49Co-2V at an order
		parameter fixed by neutron diffraction \cite{yu1999pinning}: identifying
		that slope with Eq.~\eqref{eq:mager} gives $\gw=0.90$\,mJ/m$^2$ and hence
		$K_1=\gw^2/16A=2.2\times10^{3}$\,J/m$^3$, in agreement with
		torque magnetometry on ternary Fe--Co
		($\simeq1.5\times10^{3}$\,J/m$^3$ \cite{major1988high}). The same source
		brackets the disordered wall energy at 1.9--3.9\,mJ/m$^2$ from ordered
		and disordered specimens compared at matched grain size, which contains
		the $S\!=\!0$ entry listed here. Only the $S\!=\!0$ row enters
		Fig.~\ref{fig:map}, since as-solidified AM Fe--Co prints under-ordered;
		the consequences of the ordered row for annealed material are part of a future work.
		$K_1$ for the Fe--Si series is from the same source; $\Js$ and $A$ are from
		\cite{coey2010magnetism} for Fe, Ni and Ni$_{80}$Fe$_{20}$ and from
		\cite{o2005modern} for the alloys, and that source cautions that $A$
		is not measured directly and that different derivations disagree, which is
		the origin of the $\pm30\%$ propagated in \ref{app:B}. The Fe-6.9Si row is
		extrapolated from the 6.37\,wt\% Si crystal of \cite{hall1959single}, the
		highest Si content measured there.}
	\label{tab:const}
	\resizebox{\textwidth}{!}{%
	\begin{tabular}{lccccccc}
		\toprule
		Alloy & $\Js$ (T) & $K_1$ (J/m$^3$) & $A$ (J/m) & $\tilde\lambda$ &
		$\dw$ (nm) & $\gw$ (mJ/m$^2$) & Source\\
		\midrule
		pure Fe      & 2.15 & $4.6\times10^4$ & $2.2\times10^{-11}$ & $1.57\times10^{-5}$ & 69 & 4.02 & a,c\\
		Fe-3Si       & 2.01 & $3.65\times10^4$ & $1.8\times10^{-11}$ & $4.1\times10^{-6}$ & 70 & 3.24 & a,b\\
		Fe-6.9Si     & 1.80 & $2.0\times10^4$ & $1.7\times10^{-11}$ & $2.0\times10^{-6}$ & 92 & 2.33 & a,b\\
		Fe-49Co-2V, $S\!=\!0$ & 2.35 & $2.5\times10^4$ & $2.3\times10^{-11}$ & $2.5\times10^{-5}$ & 95 & 3.03 & b,d,f\\
		Fe-49Co-2V, $S\!=\!0.88$ & 2.35 & $2.2\times10^3$ & $2.3\times10^{-11}$ & $2.5\times10^{-5}$ & 323 & 0.90 & f,g\\
		Ni           & 0.61 & $-5.0\times10^3$ & $0.8\times10^{-11}$ & $2.4\times10^{-5}$ & 126 & 0.80 & c\\
		Ni$_{80}$Fe$_{20}$ & 1.04 & $-2.0\times10^3$ & $0.7\times10^{-11}$ & $2.0\times10^{-6}\,\dagger$ & 186 & 0.47 & c\\
		FCC FeCoNi(SiAl) HEA & 1.48 & $3\times10^3$ & $1.0\times10^{-11}$ & $3$--$10\times10^{-6}$ & 181 & 0.69 & e\\
		\bottomrule
	\end{tabular}
}
	\\[2pt]
	\footnotesize (a)~\cite[Table I]{hall1959single}; (b)~\cite{o2005modern};
	(c)~\cite[Table 11.2 and data sheets pp.~385--388]{coey2010magnetism};
	(d)~\cite{hall1960magnetic};
	(e)~\cite{song2023evolution,zhang2019compositional,han2022mechanically};
	(f)~\cite[Table 4]{sundar2005soft}; (g)~\cite{yu1999pinning}.
	Composition-tuned commercial permalloy is adjusted towards $K_1\to0$ and
	$\tilde\lambda\to0$, moving it further into the process-forgiving corner of
	Fig.~\ref{fig:map} than the nominal 80:20 entry listed here.
\end{table}

Table~\ref{tab:const} separates input from output: the quartet $(\Js,K_1,A,\tilde\lambda)$ is taken from the sources indicated there, while $\dw$ and $\gw$ are computed from $A$ and $K_1$ and are therefore reproducible by the reader without recourse to any further data. Where a handbook quotes a range, the midpoint is used and the range is propagated as described below.

All arithmetic below uses $\gw=4\sqrt{AK_1}$ and $\dw=\pi\sqrt{A/K_1}$ evaluated
from the single pair $(A,K_1)$ listed for each alloy in Table~\ref{tab:const}, with
$\alpha G_\mu b=0.3\times81\,\mathrm{GPa}\times0.25\,\mathrm{nm}=6.08$\,Pa.m. For pure
iron this gives $\gw=4.02$\,mJ/m$^2$ and $\dw=68.7$\,nm. Uncertainties are propagated
once, as $\pm30\%$ in $A$ (hence $\pm14\%$ in $\gw$ and $\dw$, which depend on
$\sqrt{A}$) combined with $\pm25\%$ in the metallographic intercept, giving a factor
$1.43/0.70$ on every grain-boundary term. No constant is re-adjusted between systems.

\emph{L-PBF Fe, stress-relieved.} $H_{\rm gb}=3\gw/(2\Js D)=561$\,A/m at
$D=5\,\mu$m, plus $H_p=5$\,A/m from Eq.~\eqref{eq:neel} with $d=30\,\mu$m,
$p=0.5\%$; total 566\,A/m, band 395--810\,A/m; measured 425 \cite{zanni2022relationship}.

\emph{L-PBF Fe, anneal decrement.} This row involves neither $\gw$ nor $\dw$ and is
therefore insensitive to the exchange-stiffness uncertainty. Eq.~\eqref{eq:trauble} with $c_1=0.32$ and
$\bar\rho:1.6\times10^{14}\to1.6\times10^{13}\,\mathrm{m^{-2}}$
\cite{lejvcek2019selective,song2014vacuum} gives $269-85=184$\,A/m; measured 185.
The grain-boundary and porosity terms cancel in the difference, so this row tests
$H_\rho$ alone.

\emph{High-purity Fe.} $H_{\rm gb}=28$\,A/m at $D=100\,\mu$m, band 20--40;
measured $\approx$40 \cite{degauque1982influence}, a 30\% under-prediction.

\emph{Fe-6.9Si, as-built.} $\gw=2.33$\,mJ/m$^2$, $\dw=91.6$\,nm. Mager floor
65--194\,A/m at $D=10$--30\,$\mu$m; dislocation term 85\,A/m from
$\bar\rho=2.5\times10^{14}\,\mathrm{m^{-2}}$ ($\lambda_c\simeq0.5\,\mu$m via
Eq.~\eqref{eq:holt}), $E=1.81$ and $\tilde\lambda=2.0\times10^{-6}$; total
149--279\,A/m; measured $\approx$100 \cite{garibaldi2018effect}. This is the one
row outside a factor of two; see Section~\ref{sec:val}.

\emph{Fe-6.9Si, after 1150\,$^\circ$C.} Mager floor 6.5\,A/m at $D\approx300\,\mu$m,
band 4.6--9.3. The measured 16\,A/m exceeds the floor, as it must: the excess is
residual pinning the anneal did not remove, and the row tests the floor as a bound
rather than as an equality.

\emph{HEA, as-built.} $\gw=0.693$\,mJ/m$^2$, $\dw=181$\,nm, $E=1.18$ at
$\lambda_c\simeq0.5\,\mu$m. Mager 23--70\,A/m ($D=10$--30\,$\mu$m) plus
$H_\rho=110$--366\,A/m ($\bar\rho=2.5\times10^{14}\,\mathrm{m^{-2}}$,
$\tilde\lambda=3$--$10\times10^{-6}$); total
133--436\,A/m, bracketing the measured 188 \cite{song2023evolution}.

\emph{Inclusion-pinning constants.} $c_4=0.3$ and $\beta=1/4$ throughout, both from
the ranges quoted in Section~\ref{sec:mod3} and neither adjusted between systems.
With these values Eq.~\eqref{eq:neel} returns 5\,A/m for L-PBF gas porosity
($d=30\,\mu$m, $p=0.5\%$), $\sim$$10^2$\,A/m for a $10^{-4}$ fraction of 50\,nm
oxides, and $1.7\times10^3$\,A/m for a dual-phase constitution at $\Delta\Js=1$\,T,
$d_f=1\,\mu$m and $p_f=30\%$. The spread of four orders of magnitude across these
three cases is produced entirely by the $\min/\max$ cutoffs at $\dw$, not by any
change of constants, and the pore value is independently bounded from above by
fact~(iv) of Section~\ref{sec:statevar}.

\section*{Acknowledgments}
This research was supported by the NSERC Alliance International Catalyst (ALLRP 592696-24), Canada.

\section*{CRediT authorship contribution statement}
\textbf{Dennis Boakye}: Writing – review and editing, Writing – original draft, Visualization, Validation, Methodology, Investigation, Formal analysis, Data curation. \textbf{Eric K. K. Abavare}: Writing – review and editing, investigation, conceptualization. \textbf{Chuang Deng}: Writing – review and editing, supervision, software, resources, project administration, investigation, Fund Acquisition, conceptualization.

\section*{Declarations}
The authors declare that they have no known competing financial interests or personal relationships that could have influenced the work reported in this paper.

\section*{Data availability}
The data used for the study are in the main text.

\scriptsize
\bibliographystyle{elsarticle-num} 
\biboptions{sort&compress}
\bibliography{references}

\end{document}